\documentclass[useAMS,usenatbib]{mn2e}
\usepackage{graphicx}

\title[Microlensing of an extended source]
{Microlensing of an extended source by a power-law mass
distribution}

\author[A. B. Congdon, C. R. Keeton and S. J. Osmer]
{Arthur B. Congdon$^1$, Charles R.\ Keeton$^1$ and  S.\ J.\ Osmer$^2$ \\
$^1$Department of Physics and Astronomy, Rutgers University, 136
Frelinghuysen Road, Piscataway, NJ 08854 USA\\
$^2$sjosmer@gmail.com}

\begin{document}

\maketitle

\begin{abstract}
Microlensing promises to be a powerful tool for studying distant
galaxies and quasars.  As the data and models improve, there are
systematic effects that need to be explored.  Quasar continuum and
broad-line regions may respond differently to microlensing due to
their different sizes; to understand this effect, we study
microlensing of finite sources by a mass function of stars. We find
that microlensing is insensitive to the slope of the mass function
but does depend on the mass range. For negative parity images,
diluting the stellar population with dark matter increases the
magnification dispersion for small sources and decreases it for
large sources. This implies that the quasar continuum and broad-line
regions may experience very different microlensing in
negative-parity lensed images. We confirm earlier conclusions that
the surface brightness profile and geometry of the source have
little effect on microlensing. Finally, we consider non-circular
sources.  We show that elliptical sources that are aligned with the
direction of shear have larger magnification dispersions than
sources with perpendicular alignment, an effect that becomes more
prominent as the ellipticity increases. Elongated sources can lead
to more rapid variability than circular sources, which raises the
prospect of using microlensing to probe source shape.
\end{abstract}

\begin{keywords}
gravitational lensing --- galaxies: structure --- dark matter ---
quasars: general
\end{keywords}

\section{Introduction}
\label{sec:intro}

Microlensing is an increasingly important tool
for studying small-scale structure in lens galaxies and source
quasars. In recent years, microlensing has been observed in a number
of multiply-lensed quasars \citep[e.g.,][]{Wozniak_2237,
Schechter_knots, Richards_BLR, Keeton_BLR, Paraficz_knots}.
Microlensing modeling has also been improving. For example,
\citet{Kochanek_lcurve} has introduced a sophisticated technique for
analyzing light curves. Even so, there are aspects of the models for
the lens and source that still need to be considered for
microlensing to reach its full potential. This is especially
important for quasar microlensing, where several length scales are
involved.

Different regions of a quasar emit radiation in roughly distinct
bands. For example, continuum (blackbody) radiation in the optical
and x-ray bands is emitted from the accretion disk surrounding the
supermassive black hole, while broad emission lines in the optical
and UV are thought to originate from clouds in a region outside of
and larger than the accretion disk. Radio emission comes from even
larger structures. Roughly speaking, a source can only be affected
by objects in the lens galaxy whose Einstein radii are comparable to
or larger than the source size.  Consequently, the continuum,
broad-line, and radio regions probe different structures in the lens
galaxy. Radio jets can be used to probe dark matter substructure
\citep{Metcalf_DM,Dalal_substructure} in lens galaxies while the
accretion disk \citep{Jaroszynski_disk,mortonson_source,Pooley_xray}
and broad-line region (BLR) are used for studying the stellar
component \citep{Schneider_BLR,Richards_BLR,Keeton_BLR}. In
principle, both methods can also be used to examine the light
source. This is of particular interest for the BLR whose properties
are not well understood \citep[e.g.,][]{Peterson_AGN}. In this paper
we investigate the potential of microlensing to deepen our knowledge
of the BLR and accretion disk and to determine both the abundance
and mass function of stars in lens galaxies.

Until recently, sources relevant for microlensing could not be
resolved. Therefore, many theoretical models have assumed a point
source, and have focused on determining properties of the lensing
galaxy. Such investigations have found that microlensing
magnification distributions are not very sensitive to the shape of
the microlens mass function if it spans an order of magnitude or so
in mass \citep[e.g.,][]{Wyithe_microlensing}.  The magnification
distributions do look different if there are two distinct mass
components: not just stars, but also ``dark matter'' that could come
in the form of a smooth mass component
\citep[e.g.,][]{Schechter_microlensing}, or a set of discrete
objects that are much less massive than the stars
\citep{Schechter_pointsource,Lewis_bimodal}.  We generalize the
previous studies by considering microlensing of an extended source.
The source size $R_S$ introduces a new length scale, which
heuristically divides the microlenses into two categories:
microlenses with $R_E \ga R_S$ are massive enough to be felt
individually, while microlenses with $R_E \la R_S$ act like
a smooth component.  We therefore conjecture that the finite source
size makes the magnification distributions sensitive to the
microlens mass function even when it spans just 1--1.5 orders of
magnitude. \citet{Lewis_bimodal} recently studied microlensing of an
extended source by a bimodal mass function; we now consider a
continuous mass function.

Studying microlensing of an extended source is especially
relevant for the BLR, because recent observations suggest it
has a size $R_{BLR}\sim 10^{16}$--$10^{18}$ cm
\citep{Richards_BLR,Keeton_BLR} comparable to stellar Einstein
radii.  Understanding the effects of source size should help
us probe BLR length scales more precisely.  We perhaps should
not expect to probe the BLR surface brightness distribution,
however; \citet{mortonson_source} suggest that microlensing
is not very sensitive to the source brightness profile.
Their assertion is based on simulations of circular sources
with several specific surface brightness profiles.  We perform
similar ``numerical experiments'' to consider other source
properties, notably shape and geometry.

The possibility of a non-circular source has not been considered in
previous microlensing analyses.  However, accretion disks viewed at
random inclinations would generically appear elliptical rather than
circular.  A similar situation may apply to the BLR if it has a
disky structure \citep[e.g.,][]{Murray_BLR_disk, Elvis_BLR_disk,
Richards_BLR}. We also consider annular accretion disks. Such models
are important for two reasons. First, quasar accretion disks have
inner radii defined by the innermost stable circular orbit of a
particle in motion around the central black hole. Second, typical
models (e.g., the Shakura-Sunyaev disk), emit over a wide range of
wavelengths, with each band corresponding to different annular
regions.

This paper is organized as follows.  Lens modeling and simulations
are discussed in \S \ref{sec:methods}.  Our results are presented in
\S \ref{sec:results}.  We first consider the general problem of how
source size and lens mass impact microlensing (\S \ref{sub:source
size}). In following subsections, we investigate the effects of dark
matter (\S \ref{sub:dm}), source profile (\S \ref{sub:profile}),
source ellipticity and position angle (\S \ref{sub:ellip}), and
accretion disk geometry (\S \ref{sub:geometry}).  We construct light
curves in \S\ref{sub:lcrv} to investigate variability timescales.
Our conclusions are summarized in \S \ref{sec:conclusions}.

\section{Methods}
\label{sec:methods}

We consider microlensing of an extended source
by a distribution of stars and dark matter.  We zoom in on a region
in the lens galaxy around a lensed image.  The size of the region is
chosen to satisfy two conditions. First, it must be large compared
to a typical stellar Einstein radius, which is the relevant scale
for microlensing. Second, it must be small compared to the scale of
the lens galaxy, namely the image separation.  The latter allows us
to take the mean densities of stars and dark matter to be constant.
These criteria are not very restrictive since Einstein radii are
typically on scales of microarcseconds while image separations are
on scales of arcseconds.

We describe the stellar population of the lens by a mass function
$dN/dm$, which gives the number of stars per unit mass between $m$
and $m + dm$. We use a power law of the form
\begin{equation}
  \frac{dN}{dm}\propto m^{-\alpha}\qquad(m_1 \leq m \leq m_2)\,,
  \label{eqn: Number Fraction}
\end{equation}
for some $m_1$ and $m_2$.  Rather than specifying the mass limits
$m_1$ and $m_2$ explicitly, we adopt the equivalent approach of
giving the ratio $m_1/m_2$ along with the mean mass $\bar{m}$. A
typical choice for the power law index is $\alpha=2.35$, the
Salpeter initial mass function. The mass function is normalized so
that the scaled mass density is $\kappa_*$. In addition to stars,
the galaxy may include a continuous component with density
$\kappa_c$. The final ingredient is the shear $\gamma$, which
accounts for tidal forces from outside the patch of stars being
considered.  To obtain a relation between $\kappa$ and $\gamma$ we
must choose a mass model for the lens galaxy.  We use a singular
isothermal ellipsoid for which $\kappa=\gamma$. This model is simple
and provides a reasonable fit to observed systems
\citep[e.g.,][]{Treu_SIE1,Rusin_SIE,Treu_SIE2}.

In the absence of microlensing, the magnification is given by
\begin{equation}
    \mu_0= [(1-\kappa)^2 - \gamma^2]^{-1}\,,
\end{equation}
where $\kappa=\kappa_*+\kappa_c$. We consider a typical bright image
with $\mu_0=\pm10$ corresponding to $\kappa=\gamma=0.5\mp0.05$.  In
microlensing, the spatial distribution of stars is random, so the
magnification at a given time will be drawn from a probability
distribution. If the distribution is broad, then chances are high
that the magnification will differ from that predicted for a smooth
model, viz. $\mu_0$. The effects of microlensing are therefore
naturally described by the dispersion (standard deviation) of the
probability distribution.

\begin{figure*}
\begin{center}
\includegraphics[width=0.95\textwidth]{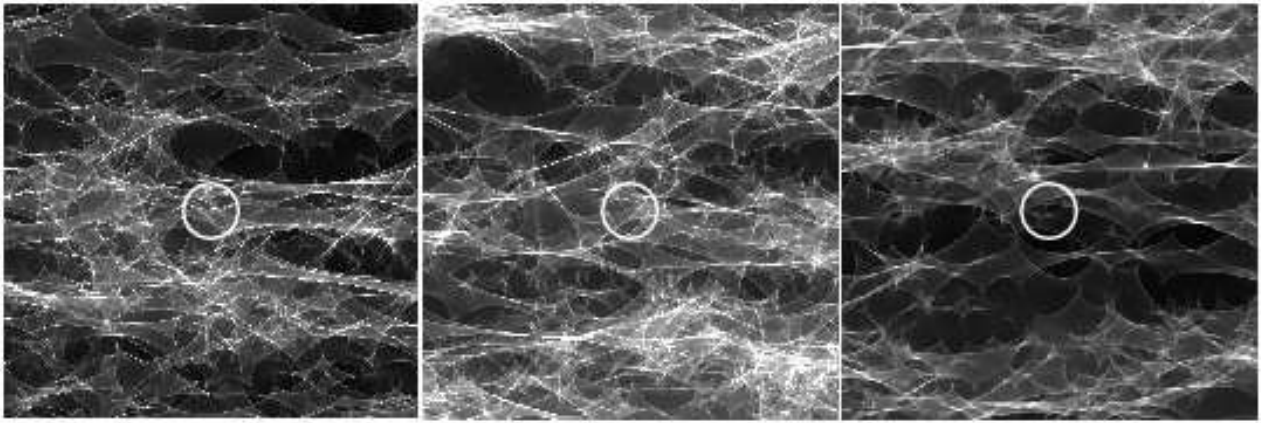}
\end{center}
\caption{Magnification maps for a positive-parity image with $\kappa
= \gamma = 0.45$, implying a macro-magnification of $\mu_0=10$.
 The gray scale indicates the magnification with values in the range
 $\mu=1$ (black) and $\mu=30$ (white).  Panels show Salpeter mass functions
 with $m_1/m_2=1$ (left), 0.1 (middle), and
0.03 (right). Each map has a side length $L=15 R_E(\bar{m})$.
Magnification histograms are generated by convolving the surface
brightness profile of the source (indicated by circles) with the
magnification map. \label{fig:magmap_imf}}
\end{figure*}

Computing the magnification distribution for given $\kappa$,
$\gamma$ and $dN/dm$ can only be done numerically. We use the
microlensing code of \citet{Wambsganss_software}, which gives
the magnification of a point source\footnote{Strictly speaking,
the map gives the magnification of a source with the shape of the
pixel.  In practice, the source sizes that interest us are much
larger than the pixels.} as a function of position over a square
region of the source plane with side length $L=15R_E(M_{\odot})$.
In mean mass units the side length is given by $L=15R_E(\bar{m})$
for $m_1/m_2=1$, $L=31.71R_E(\bar{m})$ for $m_1/m_2=0.1$, and
$L=52.26R_E(\bar{m})$ for $m_1/m_2=0.03$. We create magnification
maps with a resolution of $L/1024$ (see Figure \ref{fig:magmap_imf}).
To obtain a statistical sample, we perform 100 realizations for
each set of parameters we consider.

We must convolve the magnification map with a surface brightness
profile in order to compute the magnification of an extended source.
 We use Gaussian, uniform and linear profiles, respectively defined
by
\begin{eqnarray}
I_1(R)\equiv\frac{\ln 2}{\pi R_S^2} \exp\left(-\frac{R^2 \ln
2}{R_S^2}\right) && (0\le R< \infty), \label{eqn:src_gau}
\end{eqnarray}
\begin{eqnarray}
I_2(R)\equiv\frac{1}{2\pi R_S^2} && (0\le R \le \sqrt{2} R_S)
\label{eqn:src_uni}
\end{eqnarray}
and
\begin{eqnarray}
I_3(R)\equiv\frac{3}{4\pi R_S^2}\left(1-\frac{R}{2R_S} \right) &&
(0\le R \le 2 R_S)\,, \label{eqn:src_lin}
\end{eqnarray}
where $R_S$ is the half-light radius of the source, and the sources
are normalized to unit flux. In microlensing, the natural length
scale is the Einstein radius of the mean mass star
\citep[e.g.,][]{Lewis_sfunc}. We henceforth quote the source size as
$r_S \equiv R_S/R_E(\bar m)$.

These models are simple but nevertheless useful.  The Gaussian model
is popular in microlensing studies
\citep[e.g.,][]{Wambsganss_gau,Wyithe_gau}, so it is good to
include. The uniform disk is the simplest model conceivable, while
the linear disk has at least one physical connection.  These same
three models were used by \citet{mortonson_source}, so we can
compare their results with ours. While \citet{mortonson_source} (and
many others) applied the models to accretion disks, we imagine that
they are useful for describing BLRs as well. In particular, the
linear disk is similar to the biconical BLR of \citet{Abajas_BLR}.

We consider an elliptical source by making the substitution
\begin{equation}
I(R)\rightarrow I(\rho)/q, \label{eqn:src_ellip}
\end{equation}
with minor-to-major axis ratio $q$ and elliptical radius defined by
$\rho^2\equiv R^2 \cos^2\theta +R^2 \sin^2 \theta/q^2$. We also
consider a uniform annular disk, with inner-to-outer radius ratio
$Q$, by making the replacement
\begin{eqnarray}
I_2(R)\rightarrow I_2(R)\frac{1+Q^2}{1-Q^2}, \label{eqn:src_annul}
\end{eqnarray}
for radii satisfying
\begin{equation}
\frac{\sqrt{2}QR_S}{1+Q^2}\le R\le
\frac{\sqrt{2}R_S}{1+Q^2}.\label{eqn:annul_rad}
\end{equation}

In the following section our fiducial model assumes a Gaussian
source and a lens described by a stellar population whose masses are
given by a Salpeter mass function with $m_1/m_2=0.1$.  We assume
that $\kappa=\kappa_*=0.5\mp0.05$ for images of positive and
negative parity, respectively.  We explicitly state when other
models are to be used.

\section{Results}
\label{sec:results}

\begin{figure}
\begin{center}
{\includegraphics[width=0.45\textwidth]{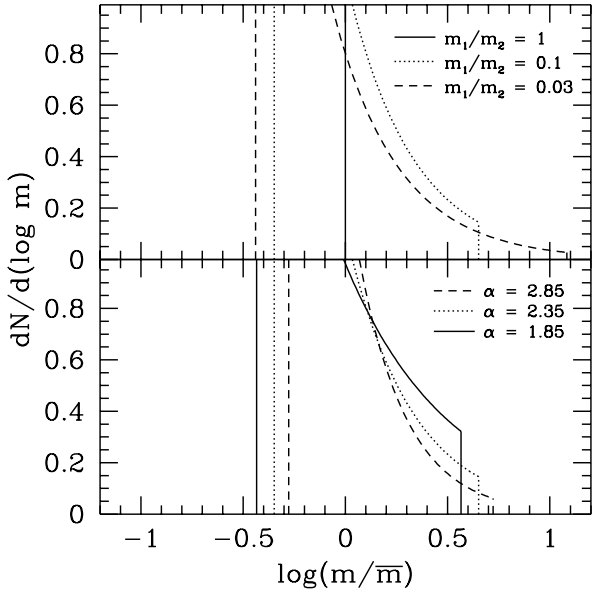}}
\end{center}
\caption{Mass functions with fixed means, $\bar{m}$, for various
mass ranges, $m_1/m_2$, and logarithmic slopes, $\alpha$. The top
panel shows mass functions with $\alpha=2.35$.  The solid, dotted
and dashed curves have $m_1/m_2$= 1, 0.1, and 0.03, respectively.
The bottom panel shows mass functions with $m_1/m_2=0.1$.  The
solid, dotted and dashed curves show $\alpha$= 1.85, 2.35 and 2.85,
respectively. \label{fig:massfunc}}
\end{figure}

\begin{figure*}
\begin{center}
\includegraphics[width=0.75\textwidth]{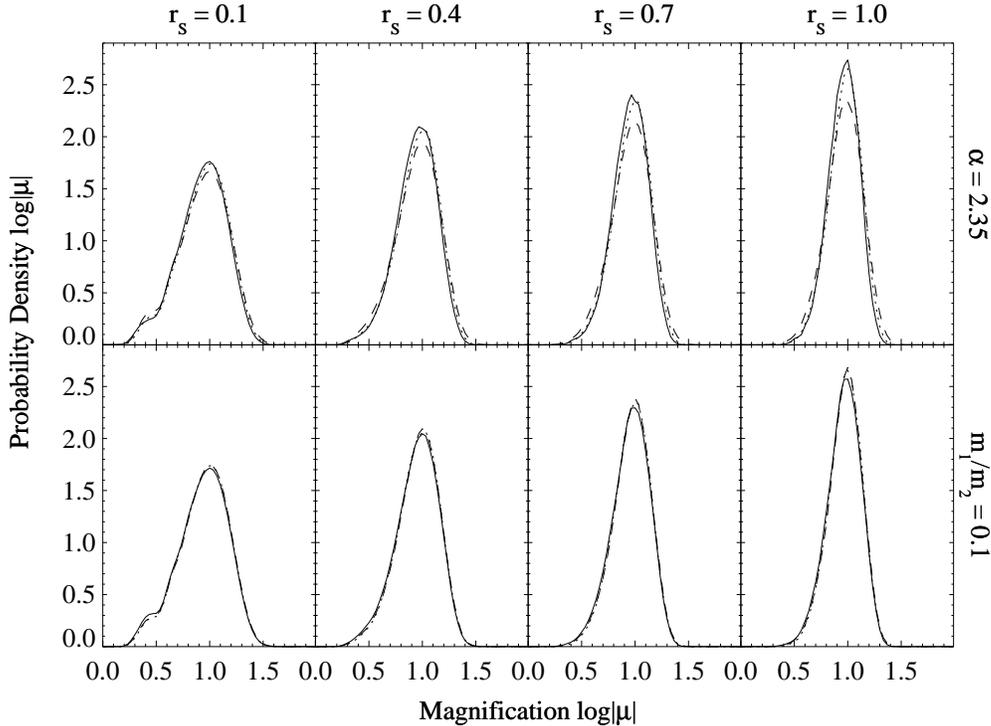}
\end{center}
\caption{Magnification histograms for different source sizes and
mass functions for a positive-parity image with
$\left|\mu_0\right|=10$. Columns show source sizes in mean-mass
Einstein radii of $r_S=$0.1, 0.4, 0.7 and 1.0. The top row shows
mass functions with logarithmic slopes of $\alpha=2.35$. The solid,
dotted, and dashed curves have $m_1/m_2$=1, 0.1 and 0.03,
respectively (see Figure \ref{fig:massfunc}, top panel). The bottom
row shows mass functions with $m_1/m_2$=0.1. The solid, dotted, and
dashed curves have $\alpha=$1.85, 2.35 and 2.85, respectively (see
Figure \ref{fig:massfunc}, bottom panel).\label{fig:hist_imf_p}}
\end{figure*}

We now study microlensing of an extended source
by a power-law mass distribution. \citet{Wyithe_microlensing}
conclude that point-source microlensing magnification distributions
are not significantly affected by the choice of mass function.  We
consider whether this result can be extended to the case of a finite
source (\S \ref{sub:source size}). Dark matter can affect
microlensing in surprising ways, raising the possibility of using
microlensing to measure the density of dark matter at the image
positions \citep{Schechter_microlensing,Schechter_dmfrac}. We
generalize earlier work by including a mass function of stars and an
extended source (\S\ref{sub:dm}).

We also explore how varying properties of the source impacts
microlensing magnification distributions.  In \S \ref{sub:profile},
we describe a source by three surface brightness profiles.  We
broaden the discussion in \S \ref{sub:ellip} to include elliptical
sources.  Finally, we consider annular sources in \S
\ref{sub:geometry}.

\subsection{Source Size and Lens Mass}
\label{sub:source size}

\begin{figure*}
\begin{center}
\includegraphics[width=0.75\textwidth]{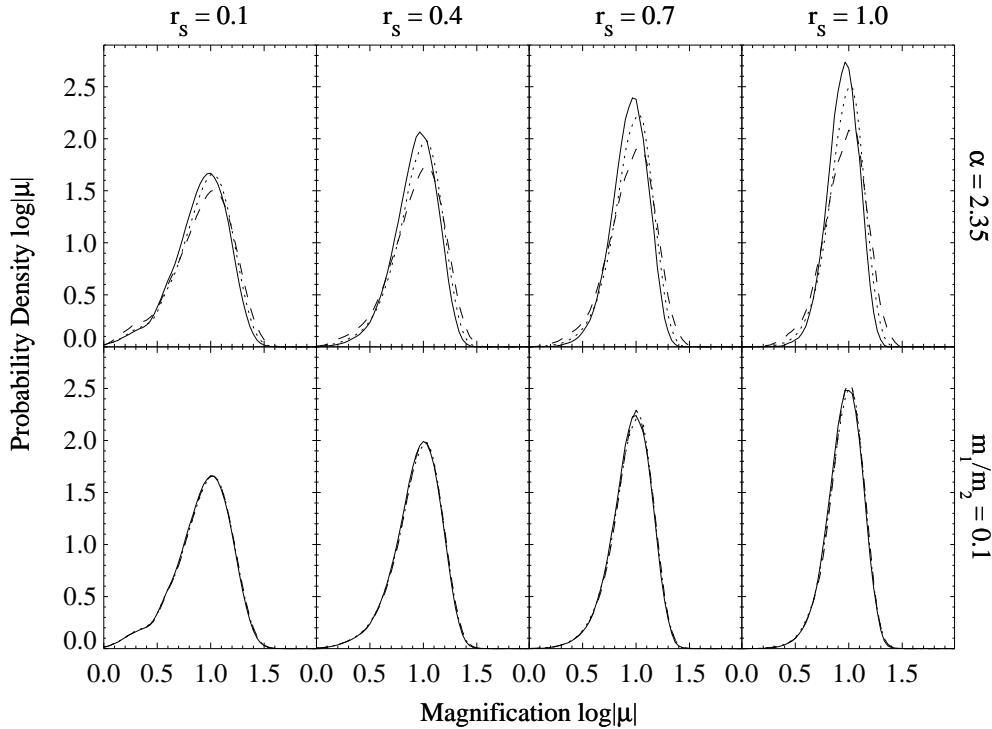}
\end{center}
\caption{Similar to Figure \ref{fig:hist_imf_p}, but for a negative
parity image with $\left|\mu_0\right|=10$. \label{fig:hist_imf_n}}
\end{figure*}

\begin{figure}
\begin{center}
\includegraphics[width=0.42\textwidth]{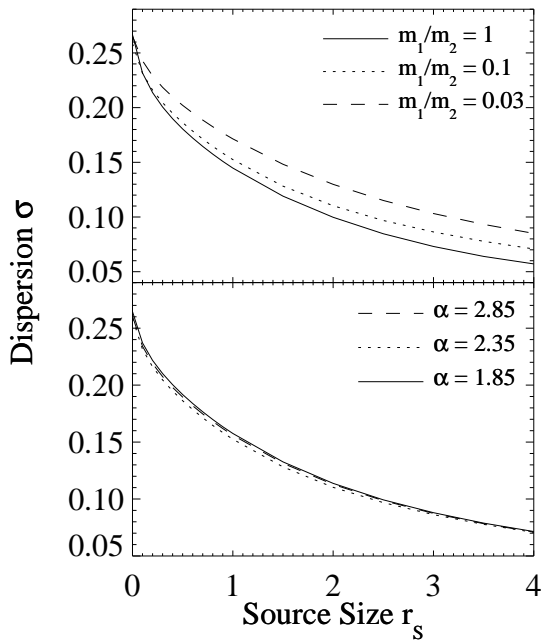}
\end{center}
\caption{Dispersion in log $|\mu|$ computed from histograms for
source sizes in the range $0 \leq r_S \leq 4$ (see, e.g., Fig.\
\ref{fig:hist_imf_p}) versus source size for different mass
functions, for a positive-parity image with $\left|\mu_0\right|=10$.
The top panel shows mass functions with logarithmic slopes of
$\alpha=2.35$. The solid, dotted, and dashed curves have $m_1/m_2$ =
1, 0.1 and 0.03, respectively. The bottom panel shows mass functions
with $m_1/m_2=0.1$. The solid, dotted, and dashed curves have
$\alpha$ = 1.85, 2.35 and 2.85, respectively. \label{fig:disp_imf}}
\end{figure}

We begin by examining how microlensing of a
finite source is affected when we vary the mass range and
logarithmic slope of the mass function
\citep[cf.][]{Wambsganss_massfunc}. Figure \ref{fig:massfunc} shows
the mass functions we use. Figures \ref{fig:hist_imf_p} and
\ref{fig:hist_imf_n} show magnification histograms for the different
mass functions and different source sizes, for positive and negative
parity.  First consider the effects of the mass range, shown in the
top row of each figure.  Increasing the mass range causes the
magnification distribution to broaden slightly, especially for
larger sources. This is shown more directly in the top panel of
Figure \ref{fig:disp_imf}, which plots the magnification dispersion
versus source size for the different mass ranges (for the positive
parity case).  When the source is small we recover the previous
result that the mass range does not affect the magnification
distribution \citep{Wyithe_microlensing}. However, as the source
gets larger there is more difference between the three mass
functions.

To understand why the magnification dispersion {\em increases} as
the mass range increases, we return to Figure \ref{fig:massfunc}.
The top panel shows that increasing the mass range causes the mass
function to ``spread out'': the lower limit decreases slightly,
while the upper limit can increase substantially.  A high upper
limit allows massive stars to exist, although they will be fairly
rare because the mass function is steep. Thus, some magnification
maps will contain one or a few massive stars that significantly
affect the microlensing, while many will not.  We believe this
explains why increasing the mass range increases the magnification
dispersion.  It also explains why the mass range becomes more
important as the source size increases:  large sources are most
sensitive to massive stars.

Now consider the slope of the mass function.  Figures
\ref{fig:hist_imf_p}--\ref{fig:disp_imf} show that the slope hardly
affects microlensing at all, regardless of the source size.  The
reason is that changing the slope shifts the mass function left or
right (see Figure \ref{fig:massfunc}), but not dramatically.

We conclude that microlensing of an extended source may offer the
possibility of determining the dynamic range ($m_1/m_2$) of the
stellar mass function, but not for determining the mass function
slope.  In light of this result, we henceforth restrict our
attention to a Salpeter mass function ($\alpha=2.35$), and we focus
attention on the case $m_1/m_2=0.1$.

\subsection{Dark Matter Content}
\label{sub:dm}

\begin{figure}
\begin{center}
\includegraphics[width=0.45\textwidth]{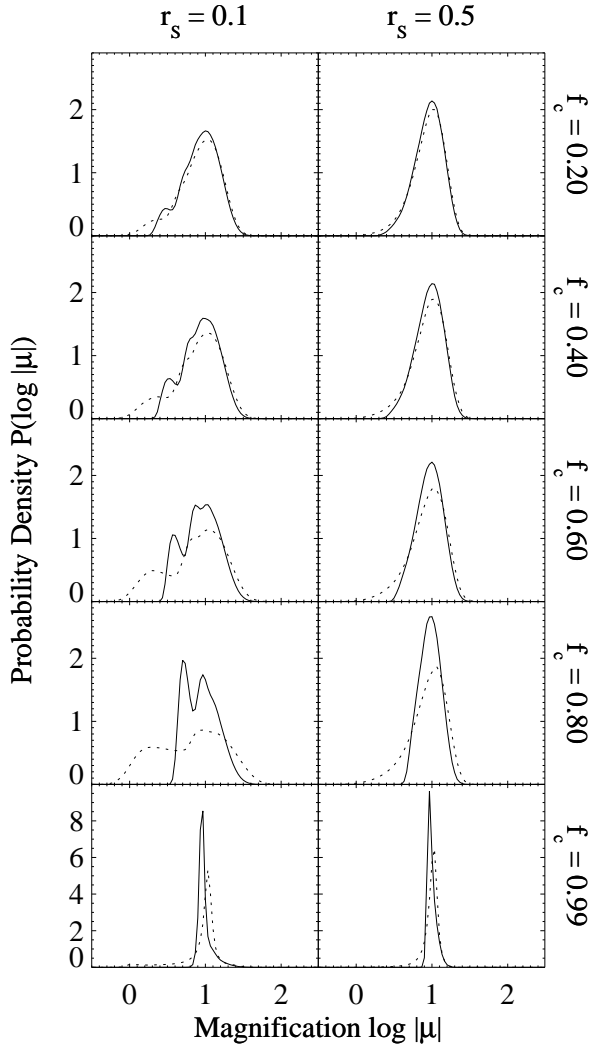}
\end{center}
\caption{Magnification histograms for an image with $|\mu_0|=10$ for
different dark matter mass fractions.  From top to bottom, panels
show $f_{\rm c}$ = 0.2, 0.4, 0.6, 0.8, and 0.99.  The left and right
columns have $r_S$ = 0.1, and 0.5, respectively.  The solid and
dotted curves have positive and negative parity, respectively. We
use a Salpeter mass function ($\alpha=2.35$) with $m_1/m_2=0.1$.
\label{fig:hist_CDM}}
\end{figure}

\begin{figure}
\begin{center}
\includegraphics[width=0.45\textwidth]{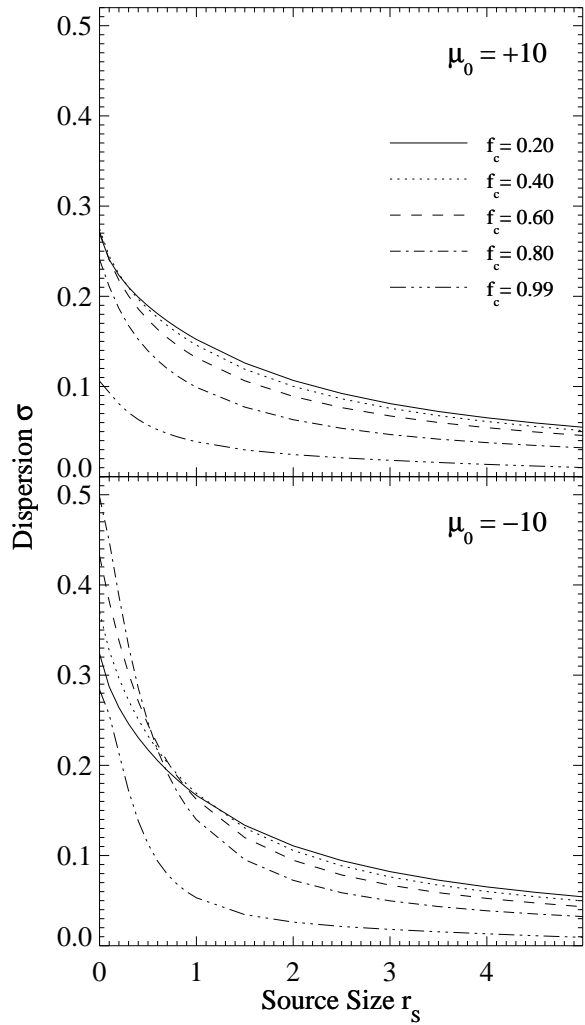}
\end{center}
\caption{Dispersion in log $|\mu|$ computed from histograms for
source sizes in the range $0 \leq r_S \leq 5$ (see, e.g., Fig.\
\ref{fig:hist_CDM}) versus source size for different dark matter
fractions.  The top (bottom) panel shows an image with positive
(negative) parity.  The solid, dotted, dashed, dot-dashed and
dot-dot-dashed curves have $f_{\rm c}$ = 0.2, 0.4, 0.6, 0.8, and
0.99, respectively.\label{fig:disp_CDM}}
\end{figure}

We now consider how the mass fraction in dark matter
affects microlensing.  Controversy remains as to whether
cosmological simulations of dark matter agree with galaxy
observations \citep[e.g.,][and references therein]{Spekkens_dm_obs,
Gerhard_ellip_dm}. Part of the problem is that most observations
(galaxy dynamics, gravitational macrolensing) depend on the global
mass distribution in a galaxy, rather than the local mass density.
\citet{Schechter_microlensing,Schechter_dmfrac} argue that
microlensing offers a new way to measure the density of dark matter
at the positions of the lensed images. They consider a uniform mass
function and a point source; we generalize the analysis to a mass
function and an extended source.

Figure \ref{fig:hist_CDM} shows magnification histograms for dark
matter mass fractions of $f_c\equiv\kappa_c/\kappa$ = 0.2, 0.4, 0.6,
0.8, and 0.99. One might expect that as $\kappa_c$ is increased,
microlensing would become less important, since fewer stars produce
simpler caustic networks. For $\kappa_c\rightarrow \kappa$ the
magnification distribution indeed approaches a $\delta$-function
centered at $\mu_0$, as seen in the bottom row of Figure
\ref{fig:hist_CDM}.

For smaller values of $\kappa_c$, however, the histograms show more
structure.  In particular, secondary peaks appear in several of the
histograms \citep[see, e.g.,][]{Rauch_bump}. The different peaks
correspond to different numbers of microimages
\citep[see][especially Figure 4]{Granot_imagenumber}. If dark matter
is the primary mass component, the probability that a source will
have multiple microimages is low. In that case, the magnification
distribution has a single peak near the expected magnification in
the absence of microlensing (e.g., $f_c=0.99$). For smaller values
of $f_c$, the density of caustics increases, which in turn raises
the probability that a source will have extra microimage pairs. The
magnification distribution therefore acquires a second peak
associated with regions of the source plane for which an extra image
pair is created (see left-hand column of Figure \ref{fig:hist_CDM}).
When $f_c$ is low, regions with extra microimages become the norm,
and the peaks corresponding to different numbers of microimages
become less distinct.  Also, an extended source often covers regions
with different numbers of microimages, smearing out the effects of
additional microimages (see right-hand column of Figure
\ref{fig:hist_CDM}).

Figure \ref{fig:hist_CDM} also demonstrates the importance of
parity. We find that negative parity images lead to distributions
with tails at low magnification. \citet{Schechter_microlensing} find
the same behaviour for a point source. As the source size is
increased, the tails become less apparent. When $r_S=1.0$ (not
shown), the two parities give nearly identical results. One surprise
is that a difference between positive and negative parity can be
observed in the skewness of the distributions even for $f_c=0.99$.
This means that even a small stellar mass fraction gives rise to
noticeable parity-dependent effects.

Figure \ref{fig:disp_CDM} uses the magnification dispersion to
quantify the effects of parity and source size.  In the positive
parity case (top panel), replacing stars with dark matter decreases
the dispersion for all source sizes, which makes intuitive sense.
However, in the negative parity case (bottom panel) when the source
is small, diluting the stars with dark matter {\em increases} the
dispersion, at least in the range $f_c \le 0.8$.
\citet{Schechter_microlensing} first found this result for a point
source and a uniform stellar mass function. We now see that it holds
for small extended sources as well.  We discover though, that when
the source is large, the trend reverses: increasing the dark matter
fraction decreases the magnification dispersion.  It seems notable
that the curves of dispersion versus source size for different dark
matter fractions all cross at roughly the same source size ($r_S
\sim 0.8$), although we do not know whether this is significant.

It is worth pointing out that \citet{Dobler_microlensing} examine
the effects of dark matter and source size on the magnification for
demagnified lensed images. They find that increasing the dark matter
fraction always decreases the magnification dispersion for a
demagnified negative parity image. (Recall that our negative parity
image is magnified.) However, for a demagnified {\em central} image,
diluting the stars with dark matter increases the dispersion for a
small source, but decreases the dispersion for a large source.
Direct comparison between those results and ours is not possible due
to the different macro parameters.  Nevertheless, it seems clear
that the effects of dark matter on microlensing depend in a
complicated way on the parity, the macro-magnification, and the
source size.

Our findings imply that the continuum and broad-line regions may
experience very different microlensing in negative-parity lensed
images.  The continuum emission region is small and should therefore
have a magnification dispersion that increases with the dark matter
fraction.  By contrast, in many cases, the BLR may be large enough
that the dispersion will decrease as the dark matter fraction
increases \citep{Richards_BLR,Keeton_BLR}.  For positive parity
images, the continuum and BLR will both have a dispersion that
decreases with the dark matter fraction. This may turn out to be a
very important physical effect allowing us to probe both the dark
matter content of lens galaxies and the structure of lensed quasars.

\subsection{Source Profile}
\label{sub:profile}

\begin{figure}
\begin{center}
\includegraphics[width=0.45\textwidth]{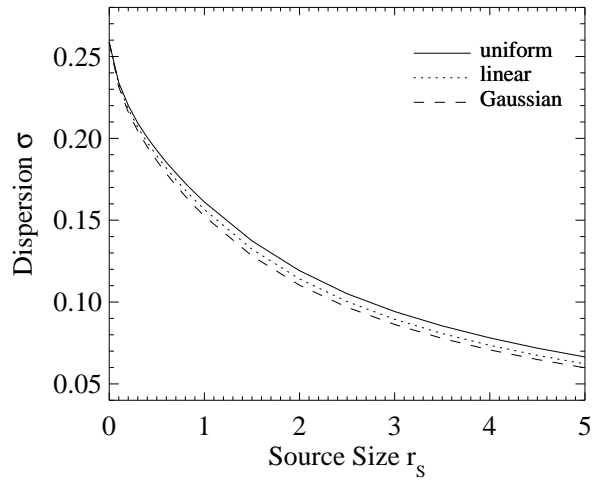}
\end{center}
\caption{Dispersion in log $|\mu|$ versus source size for a positive
parity image with different source profiles. Solid, dotted and
dashed curves have uniform, linear and Gaussian source profiles,
respectively. A Salpeter mass function with $m_1/m_2=0.1$ is used.
\label{fig:disp_srcpro}}
\end{figure}

In the remaining subsections, we return to
models consisting of a purely stellar mass component, and consider
how microlensing depends on properties of the source.  We first
examine different source surface brightness profiles, following
\citet{mortonson_source}. Figure \ref{fig:disp_srcpro} shows the
dispersion versus source size for Gaussian, uniform, and linear
profiles (defined in \S \ref{sec:methods}).  We see that the
dispersion decreases as the profile becomes steeper, although the
effect is not strong.  We therefore confirm that the dispersion
depends weakly on the source profile.

\subsection{Ellipticity and Position Angle}
\label{sub:ellip}

\begin{figure}
\begin{center}
\includegraphics[width=0.45\textwidth]{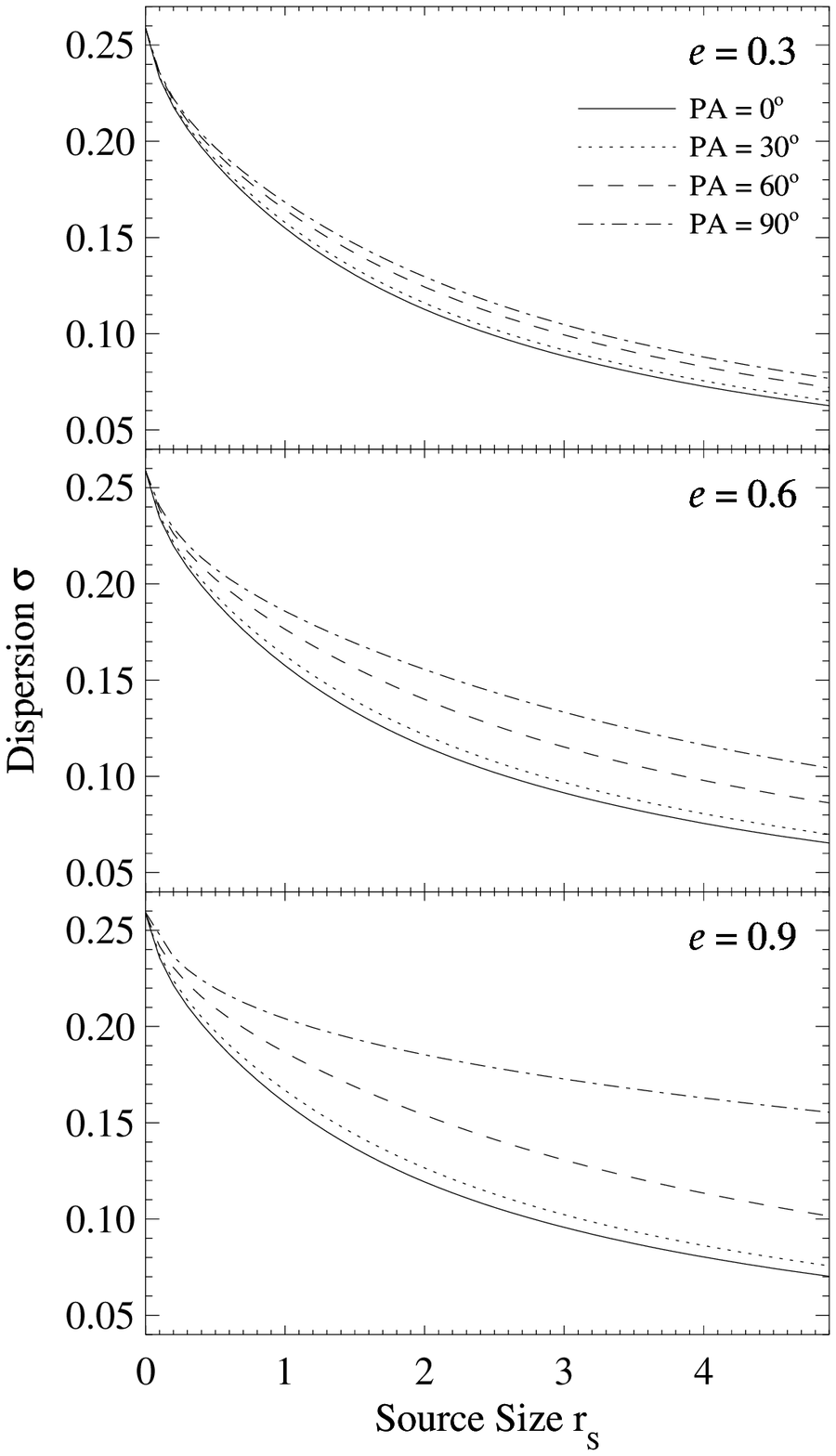}
\end{center}
\caption{Dispersion in log $|\mu|$ versus source size for a positive
parity image with different ellipticities and position angles.  From
top to bottom, panels show ellipticities of $e$=0.3, 0.6, and 0.9.
The solid, dotted, dashed and dot-dashed curves have position angles
of PA=0, 30, 60, and 90 degrees, respectively.  The source is
described by a Gaussian profile.  The stellar distribution of the
lens is modeled by a Salpeter mass function with $0.1 < m < 1$.
\label{fig:disp_ellip}}
\end{figure}

\begin{figure}
\begin{center}
{\includegraphics[width=0.42\textwidth]{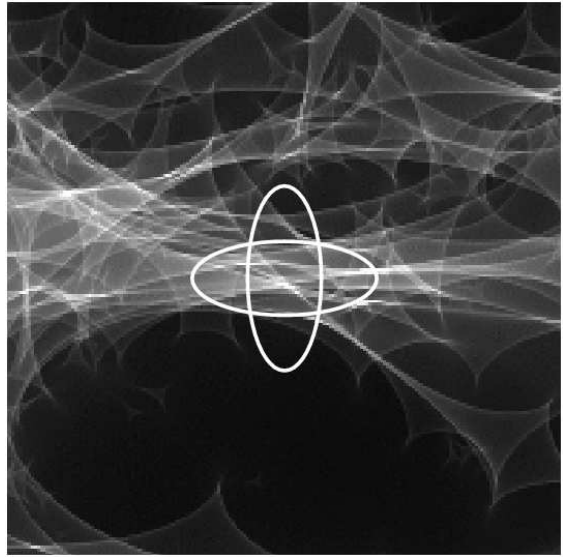}}
\end{center}
\caption{Illustration of why microlensing magnifications depend on
the orientation of an elliptical source.  Both ellipses have
semimajor axes with lengths $a = R_E(\bar{m})$ and ellipticities
$e=0.6$.\label{fig:magmap_ellip}}
\end{figure}

\begin{figure*}
\begin{center}
\includegraphics[width=0.75\textwidth]{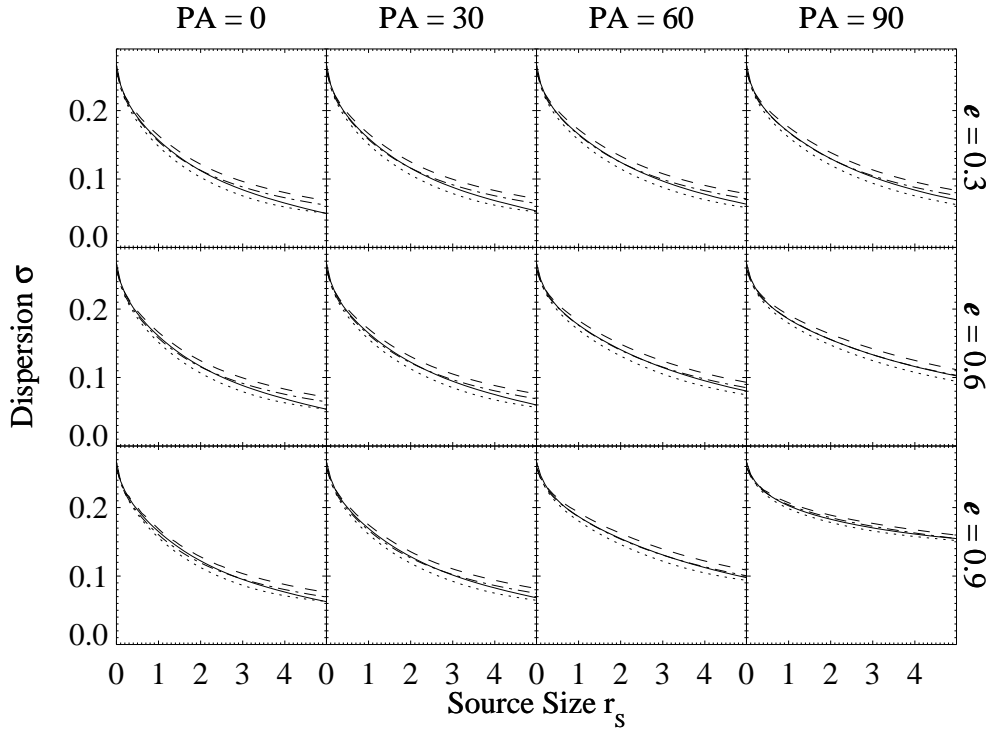}
\end{center}
\caption{Dispersion in log $|\mu|$ versus source size for a positive
parity image with different ellipticities and position angles.  Rows
show $e$=0.3, 0.6, and 0.9, while columns show PA=0, 30, 60 and 90
degrees. Solid and dashed curves have a uniform source profile.  The
solid curve shows a uniform mass function ($m_1/m_2=1$) and the
dashed curve shows a Salpeter mass function with $m_1/m_2=0.1$.
Dotted and dot-dashed curves have a Gaussian profile.  The dotted
line shows a uniform mass function, while the dot-dashed line shows
a Salpeter mass function.\label{fig:disp_ellip_pro}}
\end{figure*}

We now allow the source to be non-circular.  This
possibility has not been considered in previous microlensing
analyses, although it is an important physical effect
\citep[see][]{Kochanek_AGN}.  For a population of thin disks with
random inclinations, a face-on source is rare; the average projected
ellipticity is ${\bar e} = 0.5$. Therefore, models of microlensing
need to allow a non-circular shape for the source.  This is
certainly true for continuum microlensing, whose source is
presumably a thin accretion disk.  It may be true for broad-line
microlensing as well, given evidence that BLRs may have some disky
structure \citep[e.g.,][]{Murray_BLR_disk, Elvis_BLR_disk,
Richards_BLR}.

Figure \ref{fig:disp_ellip} shows how the magnification dispersion
depends on the ellipticity, $e\equiv 1-q$, and position angle, PA,
of the source. In each panel we see that the dispersion increases
monotonically with position angle, an effect which becomes more
pronounced for large ellipticities.  To understand this behaviour,
first note that PA=$0^\circ$ describes a source whose major axis is
orthogonal to the direction of shear, which defines the long axis of
the caustics. An extended source with PA=$0^\circ$ is likely to
cover one or more caustics regardless of where it is centered (see
Figure \ref{fig:magmap_ellip}).  Small changes in the source
position do not produce dramatic changes in the magnification.  By
contrast, for a source with PA=$90^\circ$ (aligned with the
caustics), small displacements of the source can change the number
of caustics covered, with corresponding large deviations in the
magnification. This explains why the magnification dispersion is
higher for sources aligned with the caustics (PA=$90^\circ$) than
for orthogonal sources (PA=$0^\circ$). These effects become even
more pronounced for more highly elongated sources.

In Figure \ref{fig:disp_ellip_pro} most of the effects discussed so
far are considered simultaneously.  As in Figure \ref{fig:disp_imf},
the dispersion is larger for our fiducial model ($m_1/m_2=0.1$,
dashed and dot-dashed curves) than for a uniform mass function
(solid and dotted curves). As in Figure \ref{fig:disp_srcpro}, the
dispersion is smaller for the Gaussian profile (dotted and
dot-dashed curves) than for the uniform source profile (solid and
dashed curves). Perhaps the most interesting point is that as the
source ellipticity increases, the difference between the four curves
becomes smaller, i.e., the dispersion becomes even less sensitive to
the mass function and source profile.

\subsection{Accretion Disk Geometry}
\label{sub:geometry}

Finally, we consider whether variations in
accretion disk geometry result in observable differences for
microlensing. We model the source as an annulus with a given
half-light radius, $r_S$, and hole-to-total area ratio, $Q^2$. This
approach is useful in two ways. First, quasar accretion disks have
inner radii defined by the innermost stable circular orbit of a
particle in motion around the central black hole. Second, typical
models (e.g., the Shakura-Sunyaev disk), emit over a wide range of
wavelengths. Roughly distinct annular regions within the disk are
revealed by observations in different bands \citep[see,
e.g.,][]{mortonson_source}. It is important to determine whether
microlensing can be used to find the mass of the central black hole
or the scale of the annulus within the disk emitting at some
wavelength.

\begin{figure}
\begin{center}
\includegraphics[width=0.45\textwidth]{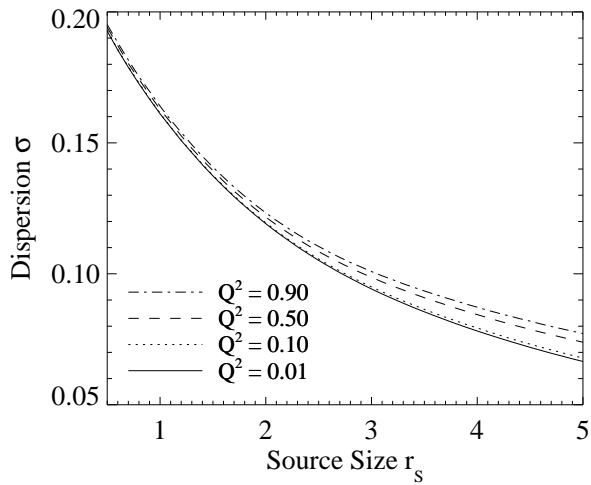}
\end{center}
\caption{Dispersion in log $|\mu|$ versus source size for a positive
parity image with different disk geometries.  The source is modeled
as an annulus with a given half-light radius.  The solid, dotted,
dashed and dot-dashed curves have hole-to-total area ratios, $Q^2=$
0.01, 0.1, 0.5 and 0.9, respectively.\label{fig:disp_annul}}
\end{figure}

For simplicity we focus on a uniform source.  Figure
\ref{fig:disp_annul} shows the dispersion versus source size for
disks with $Q^2$ = 0.01 (solid curve), 0.1 (dotted), 0.5 (dashed),
and 0.9 (dot-dashed). For small sources ($r_S\la 1.5$) the
dispersion is nearly identical for all values of $Q^2$.  For larger
sources, the dispersion remains similar for $Q^2=0.01$ and for
$Q^2=0.1$. However, the cases $Q^2=0.5$ and $Q^2=0.9$ have larger
dispersions, suggesting that only large holes can significantly
affect microlensing.

\subsection{Light Curves}
\label{sub:lcrv}

While our analysis has focused on magnification distributions,
microlensing has a time domain as well and we would like to
understand whether variability timescales can provide more
information about the lens and source.  Although a complete study of
microlensing light curves and light curve statistics is beyond the
scope of this paper, we can examine sample light curves and begin to
identify useful results. We have found that source ellipticity and
orientation have pronounced effects on the magnification
distribution, so we now see how they affect light curves.

\begin{figure}
\begin{center}
\includegraphics[width=0.45\textwidth]{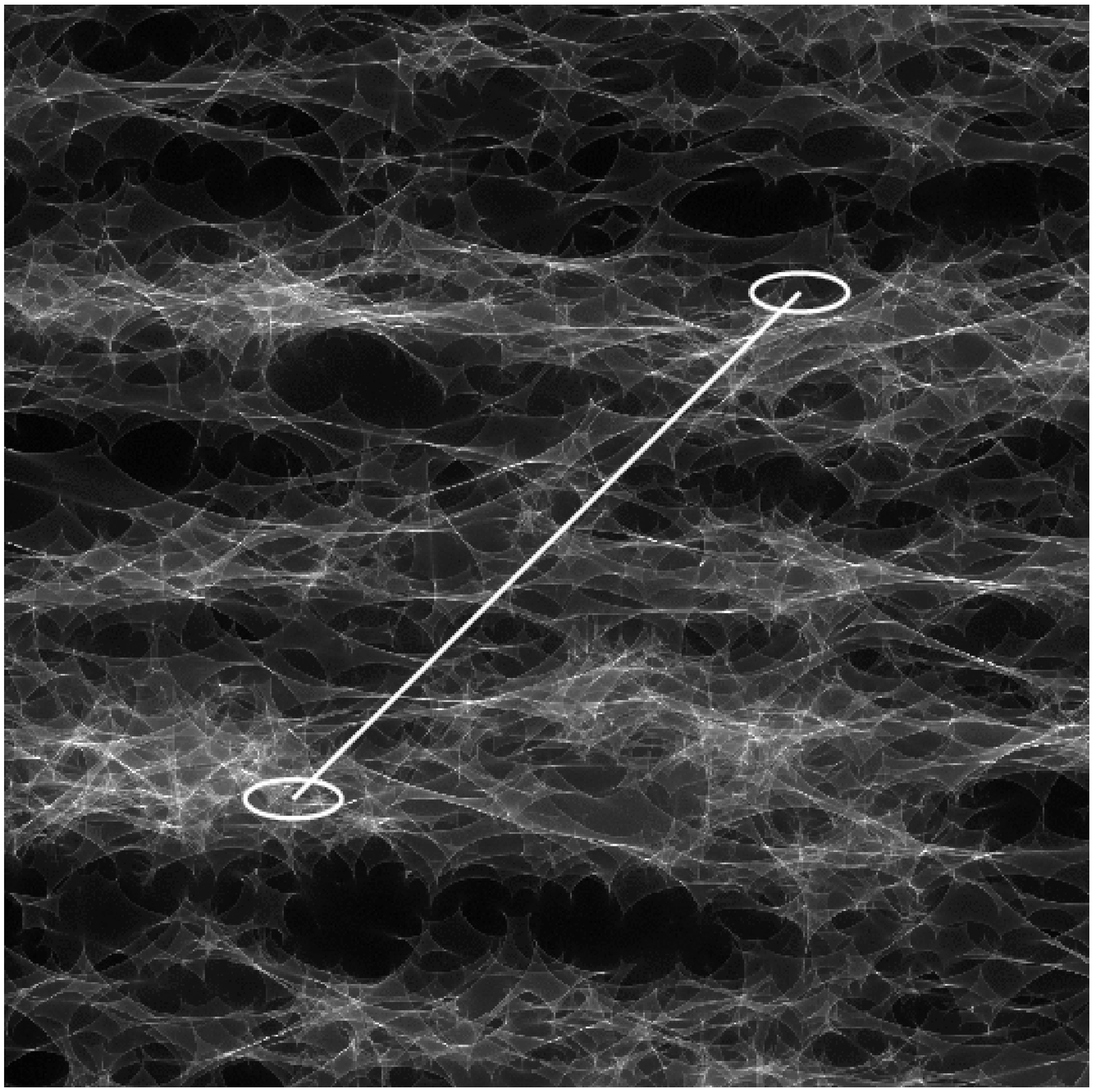}
\end{center}
\caption{Construction of light curves.  A source with ellipticity
$e=0.6$ and PA$=90^\circ$ is moved along the trajectory from lower
left to upper right.  The resulting light curves are shown in the
middle panel of the left column of Figure 14.
\label{fig:magmap_lcrv_e06}}
\end{figure}

\begin{figure*}
\centerline{
  {\includegraphics[width=0.45\textwidth]{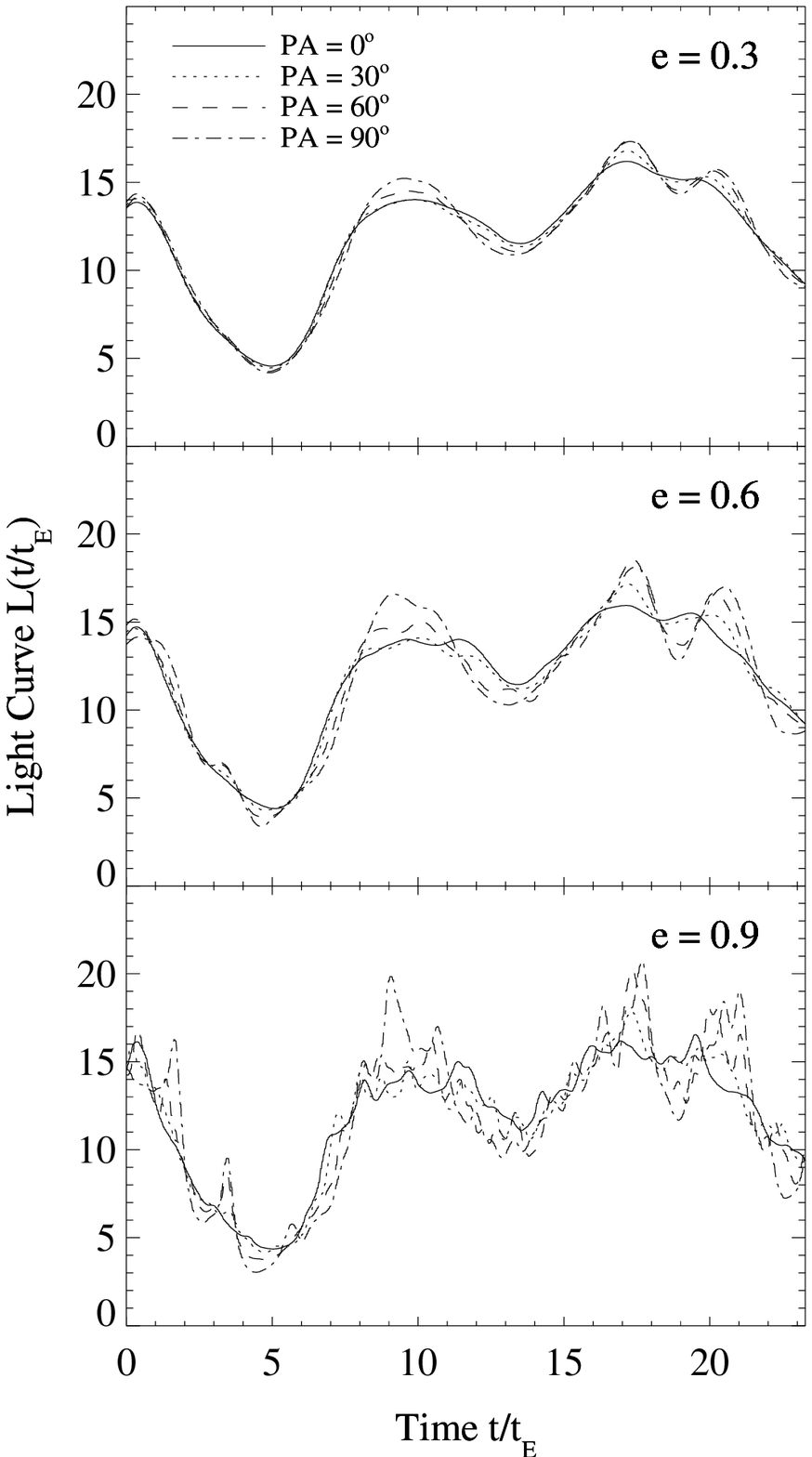}}
  {\includegraphics[width=0.45\textwidth]{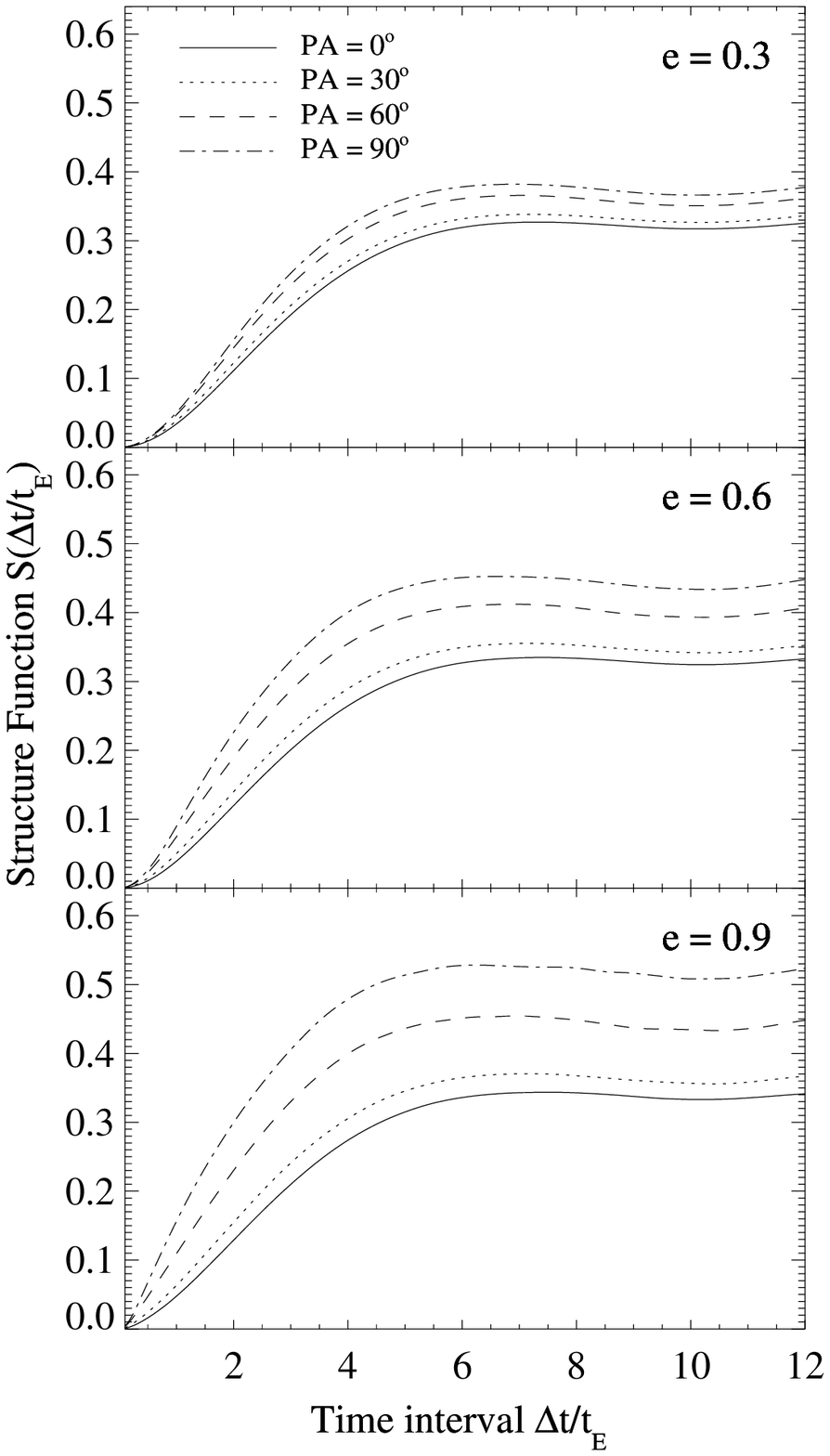}}
}
\caption{Light curves (left) and structure functions (right) for
models with $e$= 0.3, 0.6 and 0.9 (top to bottom).  Solid, dotted,
dashed and dot-dashed curves have PA = 0, 30, 60 and 90 degrees,
respectively. \label{fig:lcrv}}
\end{figure*}

To obtain sample light curves, we move the source through the
magnification map along some trajectory, as shown in Figure
\ref{fig:magmap_lcrv_e06}. The natural time scale is the Einstein
crossing time,  $t_E = R_E({\bar m})/v_\perp$, where $v_\perp$ is
the relative transverse velocity between the lens and source. Figure
\ref{fig:lcrv} (left) shows the resulting light curves for a source
with $r_S$ = 1 and the same set of ellipticities and orientations
used in Figure \ref{fig:disp_ellip}. Increasing either the
ellipticity or the position angle increases the amount of
variability, especially on short timescales.  This is consistent
with our previous interpretation: small changes to the source
position have more effect when the source is aligned with the shear
(PA=$90^\circ$) than when the source is perpendicular
(PA=$0^\circ$). This can lead to a striking amount of rapid
variability for highly flattened sources.

To quantify the amount of variability on different timescales, we
follow \citet{Lewis_sfunc} and use the structure function as a
statistical measure of temporal variability. The structure function
is defined to be the mean square change in the brightness after time
$\Delta t$: $S(\Delta t) = \langle[M(t + \Delta t) - M(t)]^2\rangle$
where $M$ is the apparent magnitude and the average is over $t$.  To
obtain statistically meaningful results, we average the structure
function over 100 realizations of light curves for a given set of
parameters.

The structure functions are shown in Figure \ref{fig:lcrv} (right).
They all have a roughly linear rise to a plateau beginning around
$\Delta t/t_E \sim 5$.  They confirm that there is more variability
on shorter timescales when the source is elongated and/or aligned
with the shear. It is customary to define a characteristic
variability time scale as the interval at which the structure
function reaches half its plateau value \citep[see][]{Lewis_sfunc,
Schechter_knots}. We see that this time scale can vary by a factor
of $\sim$2 depending on the ellipticity and orientation of the
source.

The important implication is that elongated sources can lead to more
rapid variability than circular sources that have identical
half-light radii; this effect must be taken into account when
interpreting observed variability timescales.  It is not clear
whether source shape can explain the rapid variability observed by
\citet{Schechter_knots} and \citet{Paraficz_knots}, but it should at
least be considered as a possible alternative to relativistic
motion.

\section{Conclusions}
\label{sec:conclusions}

We have presented a systematic study of
microlensing of finite sources.  We have extended earlier work by
combining an finite source with a lens described by a stellar mass
function. Following \citet{mortonson_source} we have explored how
the surface brightness profile and geometry of the source affect
microlensing (\S\ref{sub:profile} and \S\ref{sub:geometry}).  We
find that both effects are of minimal importance although subtle
differences are apparent: making the source surface brightness
profile steeper (Fig.\ref{fig:disp_srcpro}) and introducing a large
hole in the source (Fig.\ref{fig:disp_annul}) both increase the
magnification dispersion.

The mass function can play a more significant role. Although the
slope of the mass function does not lead to noticeable changes in
the magnification dispersion, the dynamic range can be important.
The dispersion for a finite source becomes larger as the mass range
increases.  This result has been seen before for a broad bimodal
mass function \citep{Schechter_pointsource,Lewis_bimodal}, but we
have demonstrated it for a continuous and relatively narrow mass
function as would be appropriate for stars.  This raises the
possibility of using microlensing to determine the dynamic range of
stellar mass functions in distant galaxies.

Our discussion of dark matter in \S\ref{sub:dm} reveals many
interesting results.  We find that the monotonic increase in
dispersion for a point source lensed by a mixture of stars and dark
matter \citep{Schechter_microlensing} extends to the case of
combining a small finite source with a power-law mass distribution.
However, for moderately large sources we find that microlensing
becomes less pronounced as the dark matter mass fraction is
increased.  As in previous studies
\citep[e.g.,][]{Schechter_microlensing}, we find multiple peaks in
the magnification histograms for moderate dark matter fractions.  We
also see that negative parity images have tails to low
magnifications \citep[see][]{Schechter_microlensing}, but only when
the source is small.

Finally, we have for the first time considered non-circular sources
with a range of position angles.  We find that sources aligned with
the shear have larger magnification dispersions than sources
orthogonal to the shear.  We suggest that an aligned source is much
more sensitive to the position relative to caustics than an
orthogonal source.

Apart from source size, which is fundamentally important, we believe
that two of the effects we have identified have important physical
implications.  First, the continuum and BLR will be very different
in their sensitivity to dark matter near a lensed image,
particularly a negative-parity image.  Thus, attempts to measure the
dark matter content of galaxies with microlensing
\citep[see][]{Schechter_dmfrac} would greatly benefit from {\em
spectroscopic} observations \citep[see][]{Keeton_BLR}. Second,
elliptical sources, which are relevant for inclined disks, may
experience much more rapid variability than circular sources. This
effect will surely be important when interpreting microlensing
variability time scales.

\section*{Acknowledgements}

We are especially grateful to Joachim Wambsganss for allowing us to
use his  inverse ray-shooting software, and for his helpful comments
on the manuscript. We also thank Greg Dobler, Jerry Sellwood and Tad
Pryor for useful discussions. ABC would like to thank Tim Jones and
Erik Nordgren for their input. ABC is supported by an NSF Graduate
Research Fellowship.


\begin{thebibliography}{}

    \bibitem[Abajas et al.(2002)]{Abajas_BLR}
    Abajas C., Mediavilla E., Mu\~{n}oz J. A., Popovic L. C., Oscoz
    A., 2002, ApJ, 576, 640

    \bibitem[Dalal \& Kochanek(2002)]{Dalal_substructure}
    Dalal N., Kochanek C. S., 2002, ApJ, 572, 25

    \bibitem[Dobler, Keeton \& Wambsganss(2006)]{Dobler_microlensing}
    Dobler G., Keeton C., Wambsganss, J., astro-ph/0507522

    \bibitem[Elvis(2000)]{Elvis_BLR_disk}
    Elvis M., 2000, ApJ, 545, 63

    \bibitem[Gerhard(2006)]{Gerhard_ellip_dm}
    Gerhard O., 2006, Planetary Nebulae Beyond the Milky Way,
    ESO Astrophysics Symposia, European Southern Observatory.
    Springer, p. 299

    \bibitem[Granot, Schechter \& Wambsganss(2003)]{Granot_imagenumber}
    Granot J., Schechter P. L., Wambsganss J.,
    2003, ApJ, 583, 575

    \bibitem[Jaroszy\'{n}ski, Wambsganss \& Paczy\'{n}ski(1992)]{Jaroszynski_disk}
    Jaroszy\'{n}ski M., Wambsganss J., Paczy\'{n}ski B., 1992, ApJ,
    396, L65

    \bibitem[Keeton et al.(2006)]{Keeton_BLR}
    Keeton C. R., Burles S., Schechter P. L., Wambsganss J.,
    2006, ApJ, 639, 1

    \bibitem[Kochanek(2004)]{Kochanek_lcurve}
    Kochanek C. S., 2004, ApJ, 605, 58

    \bibitem[Kochanek et al.(2006)]{Kochanek_AGN}
    Kochanek C. S., Dai X., Morgan C., Morgan N., Poindexter
    S., Chartas G., 2006, astro-ph/0609112

    \bibitem[Lewis \& Gil-Merino(2006)]{Lewis_bimodal}
    Lewis G. F., Gil-Merino R., 2006, ApJ, 645, 835

    \bibitem[Lewis \& Irwin(1996)]{Lewis_sfunc}
    Lewis G. F., Irwin M. J., 1996, MNRAS, 283, 225

    \bibitem[Metcalf \& Madau(2001)]{Metcalf_DM}
    Metcalf R. B., Madau P., 2001, ApJ, 563, 9

    \bibitem[Mortonson, Schechter \& Wambsganss(2005)]{mortonson_source}
    Mortonson M. J., Schechter P. L., Wambsganss J., 2005, ApJ,
    628, 594

    \bibitem[Murray \& Chiang(1998)]{Murray_BLR_disk}
    Murray N., Chiang J., 1998, ApJ, 494, 125

    \bibitem[Paraficz et al.(2006)]{Paraficz_knots}
    Paraficz D., Hjorth J., Burud I., Jakobsson P.,
    El\'{i}asd\'{o}ttir \'{A}., 2006, A\&A, 455, L1

    \bibitem[Peterson \& Horne(2005)]{Peterson_AGN}
    Peterson B. M., Horne, K., 2005, in {\it Planets to
    Cosmology: Essential Science in the Final Years of
    the Hubble Space Telescope (Space Telescope Science
    Institute Symposium Series)} eds. Casertano S., Livio, M.
    Cambridge University Press: Cambridge, p.89

    \bibitem[Pooley et al.(2006)]{Pooley_xray}
    Pooley D., Blackburne J. A., Rappaport S., Schechter P.,
    2006, astro-ph/0607655

    \bibitem[Rauch et al.(1992)]{Rauch_bump}
    Rauch K., Mao S., Paczy\'{n}ski B., Wambsganss J., 1992, ApJ
    386, 30

    \bibitem[Richards et al.(2004)]{Richards_BLR}
    Richards G. T. et al., 2004, ApJ, 610, 679

    \bibitem[Rusin \& Kochanek(2005)]{Rusin_SIE}
    Rusin D., Kochanek C.S., 2005, ApJ, 623, 666

    \bibitem[Schechter et al.(2003)]{Schechter_knots}
    Schechter P. L. et al., 2003, ApJ, 584, 657

    \bibitem[Schechter \& Wambsganss(2002)]{Schechter_microlensing}
    Schechter P. L., Wambsganss J., 2002, ApJ, 580, 685

    \bibitem[Schechter \& Wambsganss(2004)]{Schechter_dmfrac}
    Schechter P. L., Wambsganns J., 2004, IAU Symposium no. 220,
    Eds: S. D. Ryder, D. J. Pisano, M. A. Walker, and K. C. Freeman.
    San Francisco: Astron. Soc. of the Pacific, p.103

    \bibitem[Schechter, Wambsganss \& Lewis(2004)]{Schechter_pointsource}
    Schechter P. L., Wambsganss J., Lewis, G. F., 2004, ApJ, 613, 77

    \bibitem[Schneider \& Wambsganss(1990)]{Schneider_BLR}
    Schneider P., Wambsganss J., 1990, A\&A, 237, 42

    \bibitem[Shakura \& Sunyaev(1973)]{Shakura_disk}
    Shakura N. I., Sunyaev R. A. 1973, A\&A, 24, 337

    \bibitem[Spekkens, Giovanelli \& Haynes(2005)]{Spekkens_dm_obs}
    Spekkens K., Giovanelli R., Haynes M. P., 2005, AJ, 129, 2119

   \bibitem[Treu \& Koopmans(2004)]{Treu_SIE1}
   Treu T., Koopmans L.V.E., 2004, ApJ, 611, 739

   \bibitem[Treu et al.(2006)]{Treu_SIE2}
   Treu T., Koopmans L.V.E., Bolton A.S., Burles S., Moustakas
   L.A., 2006, ApJ, 640, 662

    \bibitem[Wambsganss(1992)]{Wambsganss_massfunc}
    Wambsganss J., 1992, ApJ, 386, 19

   \bibitem[Wambsganss(1999)]{Wambsganss_software}
   Wambsganss J., 1999, JCAM, 109, 353

   \bibitem[Wambganss, Paczy\'{n}ski \& Schneider(1990)]{Wambsganss_gau}
   Wambsganss J., Paczy\'{n}ski B., Schneider P., 1990, ApJ, 358,
   L33

   \bibitem[Wo\'{z}niak et al.(2000)]{Wozniak_2237}
   Wo\'{z}niak P. R., Udalski A., Szyma\'{n}ski M., Kubiak M.,
   Pietrzy\'{n}ski G., Soszy\'{n}ski I., \.{Z}ebru\'{n} K., 2000,
   ApJ, 540, L65

   \bibitem[Wyithe, Agol \& Fluke(2002)]{Wyithe_gau}
   Wyithe J. S. B., Agol E., Fluke C. J. 2002, MNRAS, 331, 1041

   \bibitem[Wyithe \& Turner(2001)]{Wyithe_microlensing}
   Wyithe J. S. B., Turner E. L., 2001, MNRAS, 320, 21

\end{thebibliography}
\end{document}